\documentclass[12pt]{article}
\usepackage{amsmath,latexsym}
\usepackage{graphicx}
\begin{document}

\setlength{\unitlength}{1mm}

 \title{Semiclassical gravitational effects around global monopole in Brans-Dicke theory}
 \author{\Large $^*F.Rahaman$ and  P.Ghosh }
\date{}
 \maketitle
 \begin{abstract}
                  In recent past, W.A.Hiscock [ Class.Quan.Grav. (1990) 7,6235 ] studied the
 semi classical gravitational effects around global monopole. He obtained the vacuum expectation
 value of the stress-energy tensor of an arbitrary collection of conformal mass less free
 quantum fields ( scalar, spinor and vectors ) in the space time of a global monopole.  With
 this stress-energy tensor, we study the semi classical gravitational effects of a global
 monopole in the context of Brans-Dicke theory of gravity.

  \end{abstract}



 \bigskip
 \medskip
  \footnotetext{ Pacs Nos : 98.90 cq, 04.20 Jb, 04.50  \\
     \mbox{} \hspace{.2in} Key words and phrases  : Semi classical approach, Brans-Dicke theory, Global monopole.\\
                              $^*$Dept. of Mathematics, Jadavpur University, Kolkata-700 032, India\\
                              E-Mail:farook\_rahaman@yahoo.com
                              }

    \mbox{} \hspace{.2in}   Topological defects could be produced at a phase transition in the
    early Universe. The study of topological defects has wide applicability in many areas of
    physics. In the cosmological arena, defects have been put forward as a possible mechanism
    for structure formation. Monopole is one of the topological defects, which arises when the
    vacuum manifold contains surfaces, which can't be shrunk continuously to a point [1]. These
    monopoles have Goldstone fields with their energy density decreasing with distance as
    $\frac{1}{r^2}$.
    They are also found to have interesting features in the sense that a monopole exerts
    practically no gravitational force on its surrounding non relativistic matter but the space
    time around it has a deficit solid angle [1]. At first, Barriola and Vilenkin (BV) [2]
    showed the existence of such a monopole solution resulting from the breaking of global
    $S0(3)$ symmetry of a triplet scalar field in a Schwarzschild
    background.\\
In 1990, Hiscock [3] obtained the vacuum expectation value of the
stress energy tensor of an
    arbitrary collection of conformal mass less free quantum fields (scalar, spinor and vectors)
    in the space-time of a static monopole. Taking these non zero vacuum expectation value of
    the stress energy tensors as a source, he solved the semi classical Einstein equations for
    the quantum perturbations ( to the first order in $\hbar$ ) of the metric.\\

    The Brans-Dicke theory of gravity is widely studied and well known alternative theory of
    Einstein gravity. This theory properly accommodates both Mach's Principle and Dirac's large
    number hypothesis. In this theory a scalar field $\phi$ is coupled to gravity with coupling
    constant $\omega$ [4].\\ It is widely accepted that gravity is
    not
    given by the Einstein's action at the sufficient high energy
    scales. The present unified theories indicates that a
    gravitational scalar field should exist besides the metric of
    the spacetime. It is believed that in the early stages of the
    Universe, the coupling to  matter of the scalar field would be
    the same order as that of the metric. Since topological
    defects are formed during phase transition, so it is not
    unnatural to study them in a scalar tensor theory of
    gravitation as that of Brans-Dicke theory.\\
    In recent, Several authors  have extended the monopole solution of BV to
   scalar tensor theory [5], Lyra geometry [6], Dilaton gravity [7],   Einstein Kalb-Ramond field [8], C-field theory [9],
    Tachyonic field [10], Einstein-Cartan theory [11], etc
    .\\
    In this letter, we would like to discuss semi classical gravitational effects around global
    monopole in Brans-Dicke theory. We have taken the same vacuum expectation value of the
    stress energy tensor as obtained by Hiscock and set equal to the Brans-Dicke field equations
    to solve for the quantum perturbations (to the first order in $\hbar$) of the
    metric.\\
    According to BV, the spacetime of global monopole may be
    written as
\begin{equation}
                ds^2 = dt^2 - dr^2-r^2 \alpha^2d\Omega_2^2
         \label{Eq6}
          \end{equation}
with, $\alpha^2 = 1 - 8\pi G \eta^2 $, where $\eta$ is the scale
of symmetry breaking.

According to classical theory, the trace of the energy stress
tensor vanishes if the Lagrangian is conformally invariant.
Whereas corresponding quantized theory it acquires a trace during
renormalization. This trace is a geometrical scalar which contains
derivatives of the metric tensor [3, 12]. The trace of the vacuum
stress energy for a conformally coupled massless free field is
given by the anomaly constructed from the curvature tensors of the
spacetime as
\begin{equation}
                T^a_a = \frac{1}{ 2880\pi^2}\left[aC_{ijkl}C^{ijkl} + b ( R_{ij}R^{ij} - \frac{1}{3}
                R^2)
         + c \Box R + d R^2 \right]\label{Eq6}
          \end{equation}
The constants a, b, c and d depends on the conformal scalar  field
under consideration and other symbols have their usual meaning as
in Riemannian geometry.

 The Brans-Dicke field equations are taken as
\begin{equation}
                R_{ab} = \frac{8\pi}{\phi}\left[T_{ab}-
                \frac{1}{2}g_{ab}\frac{(2\omega+2)}{(2\omega+3)}T \right]
                + \frac{\omega}{\phi^2}\phi_{,a}\phi_{,b} +
                \frac{\phi_{;a;b}}{\phi}
            \label{Eq2}
          \end{equation}
\begin{equation}
                \nabla^\alpha \nabla_\alpha \phi = \frac{1}{2\omega +3} 8\pi T
            \label{Eq3}
          \end{equation}
 where $\phi$ is the scalar field, $\omega$ is a dimension less coupling constant and $T$
 denotes the trace of $T_a^b$ , the energy momentum tensors of the matter
 fields.\\

       In our consideration, the vacuum expectation values of the stress tensors of the quantum
    fields can be set in to the Brans-Dicke equations (3)and(4) in the semi classical approach
    to the quantum theory of gravity:
\begin{equation}
                R_{ab} = \frac{8\pi}{\phi}\left[<T_{ab}> -
                \frac{1}{2}g_{ab}\frac{(2\omega+2)}{(2\omega+3)}< T >\right]
                + \frac{\omega}{\phi^2}\phi_{,a}\phi_{,b} +
                \frac{\phi_{;a;b}}{\phi}
            \label{Eq4}
          \end{equation}
\begin{equation}
                \nabla^\alpha \nabla_\alpha \phi = \frac{1}{2\omega +3} 8\pi <T>
            \label{Eq5}
          \end{equation}
Here $<T>$ denotes the trace of $<T_{ab}> $, the vacuum stress
energy momentum tensors.\\
Here the geometrical units are used with  $ G = c = 1 $ and $\hbar
\approx 2.612 \times 10^{-66}  cm^2 $.

At first, Hiscock has studied the quantum effects due to  monopole
background in the matter fields. Using trace anomaly given in
equation (2), one can find the vacuum expectation value of the
stress energy tensor of a conformally coupled massless scalar
field of the global monopole metric given in (1) as [3]

\begin{equation}
                <T_a^b> = \frac {\hbar}{r^4}  \left[-D,-(C+D),(C+D),(C+D)  \right]
            \label{Eq1}
          \end{equation}
 where ,

 $ C = \frac{1}{ 1440\pi^2}\left[n_{sc}  + 3 n_{sp} + 12n_{vd} -
 12n_{v\zeta} \frac{2\alpha^2 + 1}{3\alpha^2 - 1 }
         \right ] \left [ \frac {(1-\alpha^2 )(3\alpha^2 - 1)}{\alpha^4}  \right ] $

with $ n_{sc} ,  n_{sp}, n_{vd} $  and
 $ n_{v\zeta} $ are respectively , the number of scalar two
 components spinor, dimensional regularized vector zeta function
 regularized vector fields present and
D is  dimensionless constant, like C, depends on the number and
spin of the component fields present and on the metric parameter
$\alpha$ ( given in eq.(1)).\\

Now we are trying to find the first order in $\hbar$ corrections
to the BV monopole in Brans Dicke theory treating the effects of
the vacuum stress energy as small perturbations. The first order
in $\hbar $  correction to the metric may be  determined by
solving the linearized Brans Dicke equations. We assume the
correction metric of the form as

$ g_{rr} = 1 + f(r)$, $ g_{tt} = 1 + g(r) $

We also assume the correction of the Brans Dicke scalar field as

$ \phi = \phi_0 + \epsilon ( r )  $

        where $\phi_0$ is a constant which may be identified with
        $\frac{1}{G}$ when $\omega\rightarrow\infty $ ( G being
        the Newtonian gravitational constant ).\\

 Here the functions $f , g $ and $ \epsilon $ should be computed to the  first order in $ \hbar $
 .\\
 In this approximation it is easy to see that [5]
 \begin{equation}
              \frac{\phi^\prime}{\phi}=\frac{\epsilon^\prime}{\phi_0},
              \frac{\phi^{\prime\prime}}{\phi}=\frac{\epsilon^{\prime\prime}}{\phi_0},
           \label{Eq11 }
     \end{equation}
          with $ f,g , \epsilon \ll 1 $.\\
 With these approximations, field equations (5) and (6)  yield
     \begin{equation}
               \frac{1}{2}g^{\prime\prime} +\frac{g^\prime}{r}
               =\frac{8\pi \hbar }{\phi(2\omega+3)r^4}[\omega(C+2D)+(2C+3D)]
           \label{Eq12}
     \end{equation}
      \begin{equation}
              - \frac{1}{2}g^{\prime\prime} +\frac{f^\prime}{r}=-\frac{8\pi \hbar }{\phi(2\omega+3)r^4}[\omega(3C+2D)+(4C+3D)]
               + \frac{\epsilon^{\prime\prime}}{\phi_0}
           \label{Eq13}
     \end{equation}
     \begin{equation}
              f + \frac{1}{2}(f^\prime - g^\prime)r=\frac{8\pi \hbar }{\phi(2\omega+3)r^2}[\omega(C+2D)+(2C+3D)]
              + r \frac{\epsilon^\prime}{\phi_0}
           \label{Eq14}
     \end{equation}
     \begin{equation}
               \epsilon^{\prime\prime} +
               2\frac{\epsilon^\prime}{r}=\frac{8\pi \hbar C
               }{(2\omega+3)r^4}
           \label{Eq15}
     \end{equation}
      Now solving the above equations, we get the following solutions:\\
\begin{equation}
           g = \frac{8\pi \hbar}{\phi_0(2\omega+3)r^2}[\omega(C+2D)+(C+3D)]-
           \frac{k_1}{r}
           \label{Eq16}
     \end{equation}

    \begin{equation}
           f = - \frac{8\pi \hbar}{\phi_0(2\omega+3)r^2}[2D \omega+(C+3D)]-
           \frac{(k_1+\frac{2k}{\phi_0})}{r}
           \label{Eq17}
     \end{equation}

      \begin{equation}
           \epsilon = \frac{4\pi \hbar C }{(2\omega+3)r^2} -
           \frac{k}{r}
           \label{Eq18}
     \end{equation}

( where $k , k_1 $  are integration constants ).\\

Both the above constants $ k , k_1 $  will be taken as zero, as it
is impossible to dimensionally construct a value of the
constants k and $ k_1 $  which are linear in $\hbar $.\\
 It is currently known that solutions of Brans-Dicke field equations do not always go over
 general relativity solutions when $ \omega \rightarrow \infty $. However, in our semi
 classical approach, we see that in the limit $ \omega \rightarrow \infty $, our solution
 reduces to Hiscock's solution, which is given by
\begin{equation}
                ds^2 = \left[1 + 4 \pi \hbar \frac{(C+2D)}{r^2}\right]dt^2 - \left[1 -  \frac{8 \pi \hbar D}{r^2}\right]dr^2
                -r^2 \alpha^2 d\Omega_2^2
         \label{Eq19}
          \end{equation}

\pagebreak

$ $

$ $

$ $

$ $

$ $

$ $

\begin{figure}[htbp]
    \centering
        \includegraphics[scale=1.2]{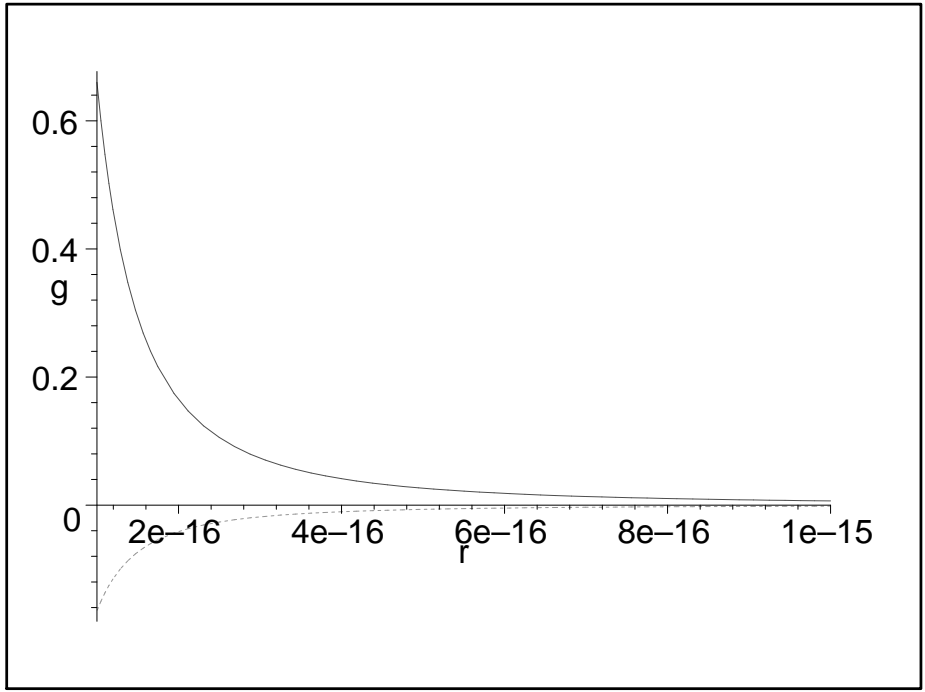}
        \caption{ We show the variation of g w.r.t. r for different values of $\omega$ (here we assume, $ \hbar = 1.05459\times10^{-34}  $ J-sec,
        $\phi_0 \sim \frac{1}{G} $, $C = 1$, $D=2$). The solid curve for $\omega = 500$ [13]  and dotted curve for
          $\omega = -1.42$ .}
    \label{fig:1}
\end{figure}

\begin{figure}[htbp]
    \centering
        \includegraphics[scale=.9]{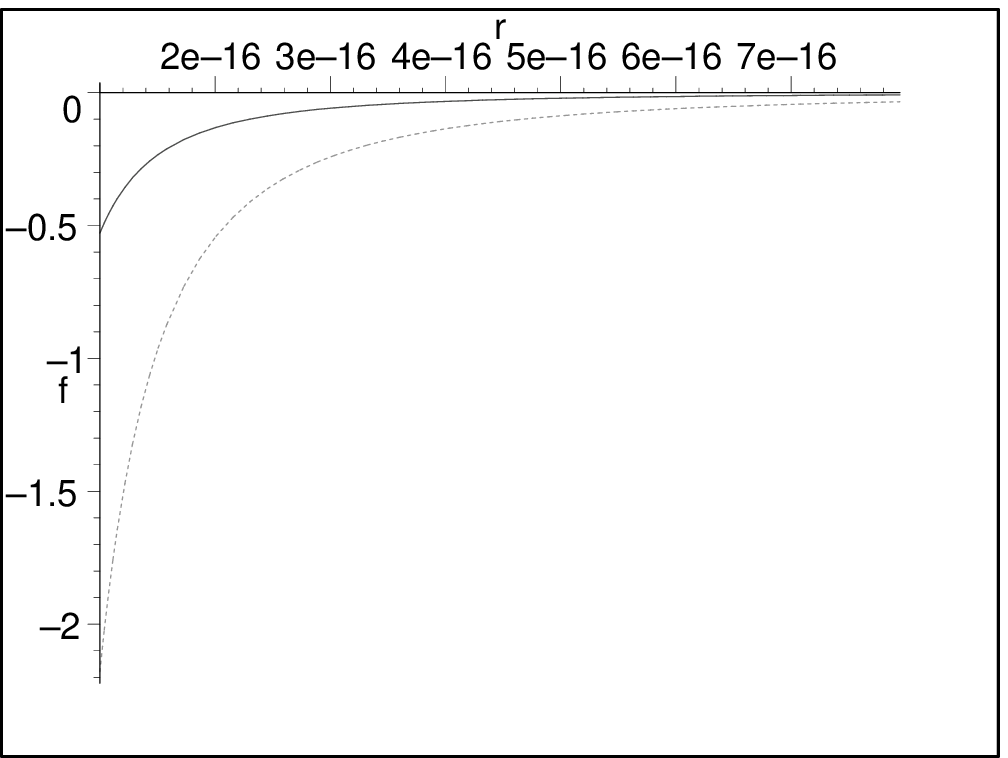}
        \caption{ We show the variation of f w.r.t. r for different values of $\omega$ (here we assume, $ \hbar = 1.05459\times10^{-34}  $ J-sec,
        $\phi_0 \sim \frac{1}{G} $, $C = 1$, $D=2$). The solid curve for $\omega = 500$ and dotted curve for
          $\omega = -1.42$ .}
    \label{fig:1}
\end{figure}

\begin{figure}[htbp]
    \centering
        \includegraphics[scale=.9]{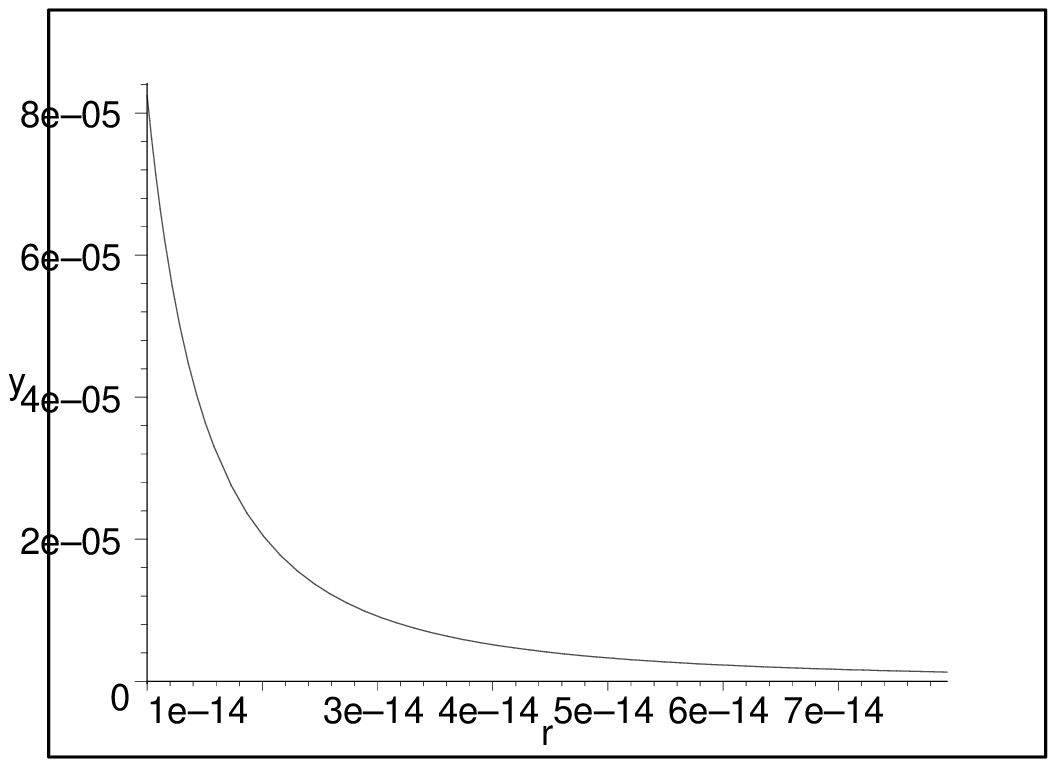}
        \caption{ We show the variation of $y\equiv \epsilon $ w.r.t. r
        . Here we assume, $ \hbar = 1.05459\times10^{-34}  $ J-sec,
        $\phi_0 \sim \frac{1}{G} $, $C = 1$, $D=2$ and
          $\omega = -1.42$ . }
    \label{fig:1}
\end{figure}

\begin{figure}[htbp]
    \centering
        \includegraphics[scale=.9]{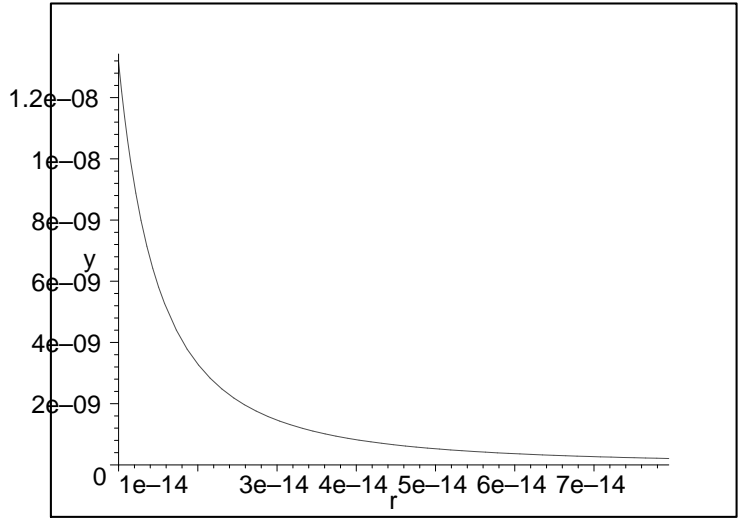}
        \caption{ We show the variation of $y\equiv \epsilon $ w.r.t. r
        . Here we assume, $ \hbar = 1.05459\times10^{-34}  $ J-sec,
        $\phi_0 \sim \frac{1}{G} $, $C = 1$, $D=2$ and $\omega = 500$. }
    \label{fig:1}
\end{figure}

\pagebreak
We shall now discuss the gravitational force of the
monopole on
the surrounding matter.\\
The radial component of the acceleration $(A^r )$ acting on a test
particle in the gravitational field of the monopole is given by
\begin{equation}
          A^r = V^r _{ ; 0}  V^0 .
         \label{Eq20}
          \end{equation}
For a co-moving particle $ V^a = \frac{1}{\sqrt{g_{00}}}
\delta_0^a $. \\
Hence using the line element (7) and solutions (19) and (20), one
can calculate $A^r$, which becomes

\begin{equation}
          A^r = \frac{- \frac{16\pi \hbar
          H}{\phi_0(2\omega+3)r^3}}{[1 +  \frac{8\pi \hbar
          H}{\phi_0(2\omega+3)r^2}]^2}
         \label{Eq21}
          \end{equation}
{where,   $ H = [ \omega(C+2D) + (C+3D)] $. }\\

Here one can see that the gravitational force varies with the
radial distance. One can note that $A^r$  may take   negative or
positive values depending on $\omega$. When, either $ \omega >Max(
- \frac{C+3D}{C+2D}, - \frac{3}{2}$) or $ \omega <Min( -
\frac{C+3D}{C+2D}, - \frac{3}{2}$), then $A^r < 0$ and this
indicates that particle accelerates towards the monopole in the
radial direction in order to keep it rest. This implies that
monopole has a repulsive influence on the test particle [14]. If,
either  $ - \frac{C+3D}{C+2D} < \omega <  - \frac{3}{2}$ or $-
\frac{3}{2} < \omega < - \frac{C+3D}{C+2D} $ , then $A^r
> 0$ and this indicates that particle accelerates away from the
monopole in the radial direction in order to keep it rest. This
implies that monopole has an attractive influence on the test
particle [14].

\begin{figure}[htbp]
    \centering
        \includegraphics[scale=.8]{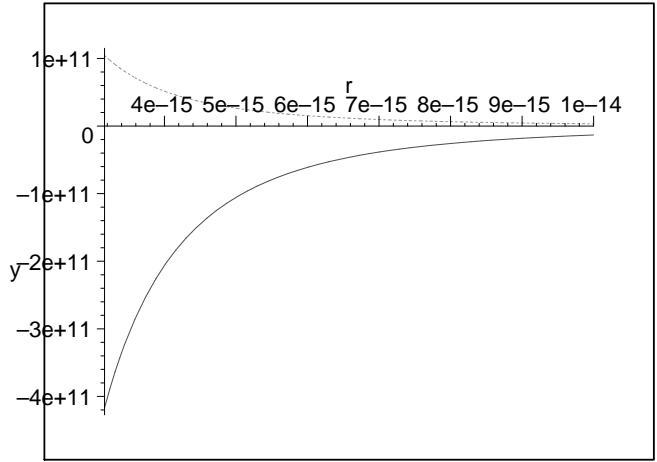}
        \caption{ We show the variation of $y\equiv A^r $ w.r.t. r for fixed $\omega$. Here we assume, $ \hbar = 1.05459\times10^{-34}  $ J-sec,
        $\phi_0 \sim \frac{1}{G} $, $C = 1$, $D=2$. The solid curve for $\omega = 500$ and dotted curve for
         $\omega = -1.42$.}
    \label{fig:1}
\end{figure}

\pagebreak
It is some interest to calculate the deficit solid
angle in the
above field.\\
The deficit solid angle $\delta$ may be defined as
\begin{equation}
          \delta(r) =4\pi [ 1 - \frac{g_{\Omega\Omega} (r)}{r^2} ]
         \label{Eq20}
          \end{equation}
in any spherical symmetric spacetime, where $R(r)$ is the proper
radius, $R = \int\sqrt{g_{rr}}dr$.
For the first order correction
to the metric (1),
\begin{equation}
          R = r + 4\pi \hbar \frac{(2\omega D + C + 3D )}{\phi_0 ( 2 \omega + 3)r}
                   \label{Eq20}
          \end{equation}
Thus in this case, we obtain,
\begin{equation}
          \delta(r) =  4\pi \left[ 1 - \alpha^2 + 8\pi \hbar \frac{(2\omega D + C + 3D )\alpha^2}{\phi_0 ( 2 \omega + 3)r^2}
          \right]
         \label{Eq20}
          \end{equation}
          Here one can  note that the deficit solid angle is a
          function of r and the first order correction to the
          deficit angle is gradually decreasing with the
          increase  of the radial distance. One can also note that
          for a fixed radial distance, the deficit solid angle decreases with the increase of $\omega > 0$
          and it is increasing with  decreasing  $\omega < -\frac{3}{2} $.  It deserves to mention
          that as $\omega \rightarrow \infty $, this deficit solid
          angle will take the same value as obtained by Hiscock in general
          relativity case.

In summary, we have studied the semi classical gravitational
effects of a global monopole in Brans-Dicke theory. We have
assumed the same vacuum expectation values of the stress energy
tensors as obtained by Hiscock and see that when $ \omega
\rightarrow \infty $, Hiscock's solution  is  recovered. We see
that our monopole exerts repulsive  gravitational force  as well
as attractive gravitational force on its surrounding non
relativistic matter depending on the Brans-Dicke coupling constant
$\omega$. This effect is absent in general relativity case
 [15]. Analogously to the general relativity case the curved
space-time of the monopole presents a deficit solid angle.

\pagebreak

        { \bf Acknowledgements: }

        F.R is thankful to DST, Government of India for providing
          financial support. We are also grateful to the anonymous referee for
        his
valuable comments and constructive suggestions. \\


\end{document}